\documentstyle[11pt,aaspp4]{article}   

\slugcomment{Submitted to ApJ}

\begin{document}

\title{The Cross-Spectra of Cir~X$-$1:
Evolution of Time Lags}
\author{J. L. Qu$^1$,  W. Yu$^1$,  and  T. P. Li$^{2,1}$}

\affil{$^1$High Energy Astrophysics Lab., Institute of High Energy Physics, CAS,
Beijing 100039, China}
\affil{$^2$Department of Physics, Tsinghua University, Beijing, China}

\begin{abstract}

Earlier work showed that the track in the X-ray
hardness-intensity diagram of Cir~X-1 corresponds to a
Z track in its color-color diagram. In this paper, we study 
the cross spectrum of Cir~X-1 in different
regions of the hardness-intensity diagram with
$RXTE$/PCA data.
Comparing the light curves of Cir~X-1 for the energy band
1.8-5.1 keV to those for 5.1-13.1 keV, we find that
Cir~X-1 exhibits a hard time lag on the horizontal branch, and a soft time
lag on both the normal and the flaring branch. This
indicates that Cir~X-1 is similar
to GX~5-1 and Cyg~X-2 on the horizontal branch, but is different
from them on the normal branch. We briefly discuss the mechanism of the time
lags in the context of Comptonization models.

\end{abstract}


\section{Introduction}

X-ray spectral and fast-timing properties show that neutron star low-mass X-ray
binaries (LMXBs) can be classified as Z sources or atoll sources (\cite{HKlis89}).
For a typical Z source, the characteristic pattern in the X-ray color-color diagram
is a Z-shaped track (hereafter Z track). The three limbs of the Z track are
called the horizontal (HB), normal
(NB), and flaring branch (FB), respectively. The transition region between the
HB  and the NB and that between the NB and the FB are called the apex (hard apex)
and the anapex (soft apex), respectively. The timing properties of a LMXB, e.g. the
characteristics of the Fourier power density spectra, are strongly correlated
with their position in the Z track (see the review by \cite{Klis95};
\cite{HKlis89}).

The Fourier cross spectrum has been used to study the frequency dependence of
the relative time lags between the intensity
variations in different energy bands. Van der Klis et al. (1987)
first used this technique to study the HB and the NB of Cyg~X-2 and GX~5-1.
They have found that the time lag of low frequency noise ( LFN ) is complex.
A soft lag, i.e., the observed soft photons lag the observed hard photons, exists
below a frequency of a few hertz on the HB. Vaughan et al.
(1994) studied the intensity variations of GX~5-1 on its HB and found
that at low frequencies the X-ray intensity variations at low energies lag those
at higher energies by tens of milliseconds.
Hard time lags have been observed in Horizontal-Branch quasi-periodic
Oscillations (HBOs) and
Normal-Branch quasi-periodic Oscillations (NBOs) in GX~5-1 ( \cite{Klis87} 1987;
\cite{Vaughan94}; \cite{Vaughan99}) and in Cyg~X-2
(\cite{Klis87} 1987; \cite{MD89}), and HBOs and Flaring-Branch quasi-periodic
Oscillations (FBOs) in Sco~X-1 (\cite{Dieters99}).

Shirey et al. ( \cite{PhD98}; \cite{Shirey98}; \cite{Shirey99} ) studied Cir~X-1
during its transition to an active state with the Rossi X-ray Timing Explorer
($RXTE$) in 1997 June. They investigated the hardness intensity diagram (HID) of
Cir~X-1 and demonstrated that the spectral branches
in the HID can be identified based on the timing properties as the horizontal,
normal, and flaring branches of a Z source, respectively.

This paper uses the cross spectrum technique to measure the time lags in Cir~X-1
and is organized as follows.
The observations and the analysis are described in Section 2, while the 
results are reported in Section 3. Our discussions and conclusions are presented in 
Section 4.

\section{The Observations and the Analysis Method}

The $RXTE$ observations of Cir~X-1 lasted 10 days from 1997 June 10 to June 20.
Based on these observations, Shirey et al. (\cite{PhD98}; \cite{Shirey98};
\cite{Shirey99}) studied the spectral and timing properties of Cir~X-1
and identified it as a Z source. The observations of Cir~X-1 on June 13
exhibited a complete track in its HID, and these are the data we use to study
the cross spectra of Cir~X-1. The data were obtained with the
Proportional Counter Array (PCA). During the observations, one of
the 5 Proportional Counter Units (PCUs) was sometimes off. For
consistency, we extract data from the same four PCUs which were on.
We only use {\sl standard 2} and {\sl single bit} modes. The {\sl standard 2}
data has 128 channels with 16 s time resolution, while the {\sl single bit} data
covers two energy bands, namely 1.8 keV - 5.1 keV (channel 0-13 )
and 5.1 keV - 13.1 keV (channel 14-35), with 122 $\mu s$ ($2^{-13}$ s) time
resolution.

We use the {\sl standard 2} data to make a HID for Cir~X-1.
Following previous studies (\cite{PhD98}; \cite{Shirey98}; \cite{Shirey99}),
we construct the HID from the background subtracted light curves in three
energy channels, namely 1.8-6.5 keV (channel 0-17), 6.5 keV-13.1 keV (channel
14-35), and 13.1-18.6 keV (channel 36-50). The result is showed in
Fig.~\ref{fig1}. To investigate the timing properties of Cir~X-1
along the Z track, we divide the HID into a number of boxes (see
Fig.~\ref{fig1}) and each box includes more than 1200 s data.

We use the {\sl single bit} data to compute power density spectra
(PDSs) and cross spectra of Cir~X-1. The data are first 
divided into 16 s segments with 4 ms time resolution and the PDSs are computed
for each segment. Then the PDSs from all the segments are averaged and 
logarithmically rebinned. The phase lags between the two energy bands 1.8-5.1
keV and 5.1-13.1 keV are quantified by means of the cross spectral analysis.
The cross spectrum is defined as
$C(j)=X^*_1(j)X_2(j)$, 
where $X_i(j)$ is the measured complex Fourier coefficient for energy band $i$
at a given frequency $f_j$. The phase lag between the signals in the two bands is
given by the Fourier phase $\phi (j)=\arg [C(j)]$ (the position
angle of the cross-vector $C$ in the complex plane). The time lag
in Fourier frequency $f_j$ is constructed from $\phi (j)$ by dividing by
$2\pi f_j$. The segments with data gaps as well as those outside boxes in
Fig.~\ref{fig1} are excluded in the analysis.

The cross spectra of Cir~X-1 on the different branches of the HID show that
the phase lags above 60 Hz are consistent with $|\phi(j)|=\pi$. This is explained
as due to the dead-time effect ( \cite{Klis87} 1987;
\cite{Vaughan99} ), which should be corrected. Thus we
subtracted a cross vector averaged over 72 to 128 Hz from the average cross
spectrum.
The corresponding white noise is subtracted from the PDS
following the model of \cite{Zhang95}.

\section{Results}

We compute the averaged cross spectra and PDSs for Cir~X-1 when it is on the
vertical HB (from box 1 to 5), on the horizontal HB (box 6-13), on the NB
(box 14-20), and on the FB (box 21-24) in the HID. The results are
shown in Fig.~\ref{fig2}A-D, respectively. The positive time lags
indicate that the observed hard photons (5.1-13.1 keV) lag the soft
photons (1.8-5.1 keV), namely hard time lags, while the negative time lags
indicate that the soft photons lag the hard photons, namely soft time lags.

Fig.~\ref{fig2}A (upper panel) indicates that on the vertical HB,
strong and narrow HBOs are present for Cir~X-1. The hard time lags
decrease with increasing Fourier frequency to about 40 Hz, and soft time lags
appear below 0.1 Hz and above about 40 Hz [Fig.~\ref{fig2}A (lower panel)]. On
the horizontal HB, a bump
near 4 Hz appears in the PDS and the centroid frequency of the
HBO increases with increasing count rate [Fig.~\ref{fig2}B (upper panel)].
On the NB, there is a wide NBO near 4 Hz in the PDS [Fig.~\ref{fig2}C (upper 
panel)], and the soft time lags
decrease with Fourier frequency [Fig.~\ref{fig2}C (lower panel)]. On
the FB, the PDS shows only VLFN [Fig.~\ref{fig2}D (upper panel)],
and the apparent negative time lags decrease slowly with frequency
[Fig.~\ref{fig2}D (lower panel)].
The time lags change from positive to negative when Cir~X-1
evolves from the vertical HB to the FB, indicating that 
its cross spectrum evolves along the track in the HID.

To investigate the correlation between the source position
in the HID and the characteristics of temporal variability, we introduce a
parameter ``$S_z$", which measures the position of Cir~X-1
on the Z track in the HID. We set $S_z$ of box 1, 6, 13, and 21
to -1, 0, 1, and 2, respectively ( see Fig.~\ref{fig1}). The $S_z$ values of
the other boxes are determined by linear interpolation.

In order to study the correlation between $S_z$ and the time lag of HBO, we
compute the average PDS in each box, then fit the PDS with a
model composed of two Lorentzians
and a power law to obtain the centroid frequency ($\nu_{HBO}$) 
and the full width at half maximum ($\Delta\nu_{HBO}$, FWHM) of the HBO. The
time lags are averaged over the frequency range
between $\nu_{HBO}-\frac{1}{2} \Delta\nu_{HBO}$
and $\nu_{HBO}+\frac{1}{2} \Delta\nu_{HBO}$. On
the horizontal branch, the Quasi-Periodic Oscillation (QPO)
fades with increasing $S_z$ and reaches
a minimum (``knee") near hard apex. So we
only compute the cross spectrum from box 1 to box 11. The average time lags
are approximately
anti-correlated with $S_z$, as shown in Fig.~\ref{fig3} . 

The NBO is present over the entire NB but is most prominently peaked
at the middle. In order to study the correlation between the time
lag near the NBO frequency range and $S_z$, we compute the average time lag 
in the range 2 - 6 Hz, and regard this as the time lag of the NBO. The correlation
is shown in Fig.~\ref{fig4}.
It shows a negative time lag that decreases with increasing $S_z$.

To summarize the results from the above studies:

\begin{itemize}
 \item The evolution of PDSs of Cir~X-1 agrees with the results by Shirey et al.
 (1999). For example, the centroid frequencies of HBOs vary from $\sim$12 Hz to
 $\sim$25 Hz on the vertical HB,  remain close to $\sim$30 Hz and fade into a
 ''knee"  on the horizontal HB, while those of NBOs peak at about 4 Hz.

\item The cross spectra of Cir~X-1 show evolution from a hard lag on the
vertical HB to a soft lag on the FB (Fig.~\ref{fig2}).

 \item The time lags in the HBO frequency range decrease with the centroid
 frequency of HBO (Fig.~\ref{fig3}).
 The time lag in the HBO frequency range changes from
 a hard lag to a soft lag near the hard apex (Fig.~\ref{fig3}).

 \item The average soft time lags between 2 Hz and 6 Hz suggest that
 the soft photons lag the hard ones in the NBO frequency range and the time
 lag shows a trend decreasing with $S_z$ (Fig.~\ref{fig4}).
\end{itemize}

The cross spectra of Cir~X-1 we showed above are similar to them of
GX~5-1 and Cyg~X-2 on the HB, but are different from those on the NB 
( \cite{Klis87} 1987; \cite{Vaughan94}). Our results also 
show that the time lags of the very low frequency noise (VLFN) on the FB are
soft time lags. 

\section{Discussion}

We have analyzed the $RXTE$ data of a complete spectral track of
the Z source Cir~X-1 during its active phase.
The time lags of the 5.1-13.1 keV photons relative
to the 1.8-5.1 keV ones are measured.
The cross spectra show that 5.1-13.1 keV photons lag 1.8-5.1 keV photons
on the HB, and that the 1.8-5.1 keV photons lag 5.1-13.1 keV photons on both the
NB and the FB. The cross spectra evolve along the track in the HID. If $S_z$,
the position of the source at the HID,
represents the mass-accretion rate of the source, the evolution of the cross
spectrum  along the track suggests that the cross spectrum varies with the
mass-accretion rate.


Both shot profile properties and Comptonization of photons can introduce time
lags. Shibazaki et al. (1988) found that energy-dependent shot profiles 
can produce low-energy time lags in the cross spectrum at frequencies of the
shot time scale (tenth of a Hertz to a few Hertz) without noticably affecting
the cross spectrum at higher frequencies. The shot model explains the cross
spectrum of GX~5-1 and Cyg~X-2 well ( \cite{Vaughan94}). However, because of the difference between the
hard lags above 0.3 Hz in the cross spectrum of 
Cir~X-1 and those of Cyg~X-2 and GX~5-1, the shot noise model may need a
modification to explain the results of Cir~X-1 in the sense that
the shot profiles do not have an obvious evolution.

On the other hand, the Comptonization models, e.g., the uniform corona model (
\cite{Payne80}), the non-uniform corona model (\cite{KHT97}), and the
drifting-blob model (\cite{BLiang99}), only explain the hard time
lags (see the review by \cite{Poutanen}). In those Comptonization
models, the photons are scattered off energetic electrons and gain energy when
they go through the hot electron clouds (corona). The observed hard photons
undergo more scatterings than the soft 
photons in the corona, therefore naturally tend to lag them. The disadvantages of
these models are that they can not explain the soft lags and a static corona
cannot be used to explain the evolution of the time lags.

In order to explain the observed soft lags and the evolution of the time lags
of QPO in the superluminal source GRS1915+105 ( \cite{Cui99}; \cite{Reig} ),
Nobili et al. (2000) proposed a Comptonization model in which the corona
consists of a hot electron cloud in the inner part and a warm one in
the outer part. If the optical depth of the hot plasma cloud is very
large, the photons that go through the inner parts of the corona can be
efficiently 
Comptonized and become harder. In the outer part, the hard photons escaping
from the inner part are scattered by warm electrons and give away their energy,
thus the soft lags would be measured.

The X-ray energy spectra below about 13 keV of Cir~X-1 observed by BeppoSAX are
well described by the Comptonization of soft photons (\cite{Iaria2000}). Suppose
that the structure of the plasma cloud in Cir~X-1 is similar to that in
GRS~1915+105, the time lags observed in Cir~X-1 can be explained by the
Comptonization model proposed by Nobili et al. (2000).

When Cir~X-1 is on the HB, the photons from the central source are
upscattered by hot electrons in the corona. They
gain energy and produce the positive
time lags. The photons may hardly undergo downscatters in the outer part which
is an optically thin corona. This leads to the hard lag.
When the mass-accretion rate increases, the electrons in the
inner part of the corona
are cooled by the photons, resulting in a smaller radius for the inner 
part of corona, thus the time lag becomes smaller. Similarly, the photons
contribute to the HBO may also go through the same process and show a decreasing
time lag with increasing $S_z$ (Fig.~\ref{fig3}).

On both the NB and the FB, the plasma in the inner part become hotter and
denser. This might be caused by an approximately radial inflow extending from
the inner accretion disk to the compact central
corona (\cite{Lamb88}; \cite{MLamb92}). The soft photons
going through the inner part are effectively Comptonized and
hardened. When they traverse to the outer cooler plasma, they
are down-scattered by electrons and give away
their energy. This introduces the soft lag.

In conclusion, the time lag observed in Cir~X-1 seems
consistent with the Comptonization model with two layers of corona.
Further study of the correlation between the Comptonization spectral component
and the time lag in Cir~X-1 is crucial to understanding the above
interpretation.

\acknowledgments

We would like to thank the anonymous referee for carefully reading this
manuscript and helpful comments. We also thank Dr. W. D. Li for kind help.
This work was subsidized by the Special Funds for Major State Basic Research
Projects and by the National Natural Science Foundation of China.

\clearpage

\figcaption[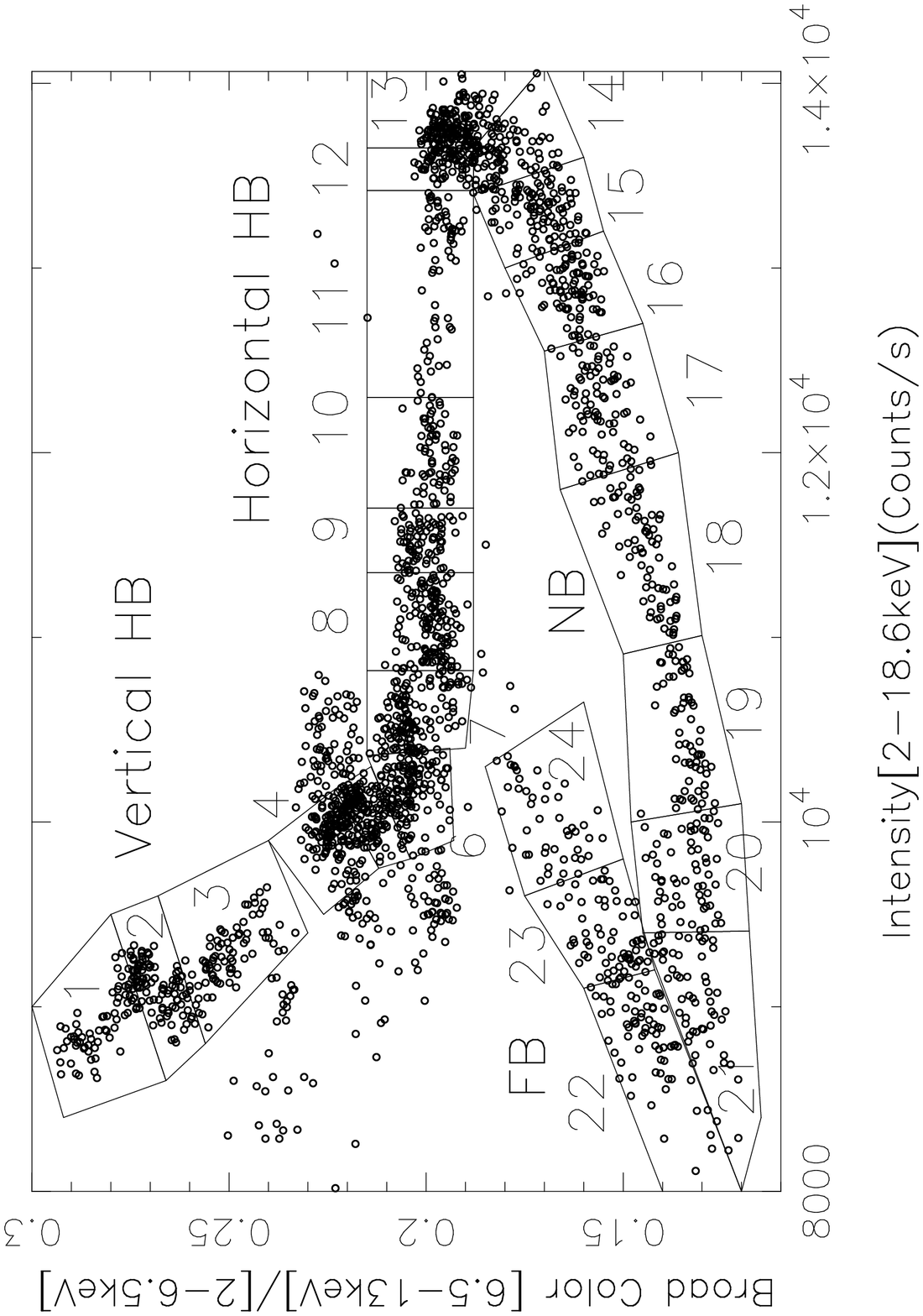]{The HID of Cir~X-1. Each point corresponds to 16 s of
background subtracted data from 4 PCUs. Each box includes $\geq$1200 s
data.\label{fig1}}

\figcaption[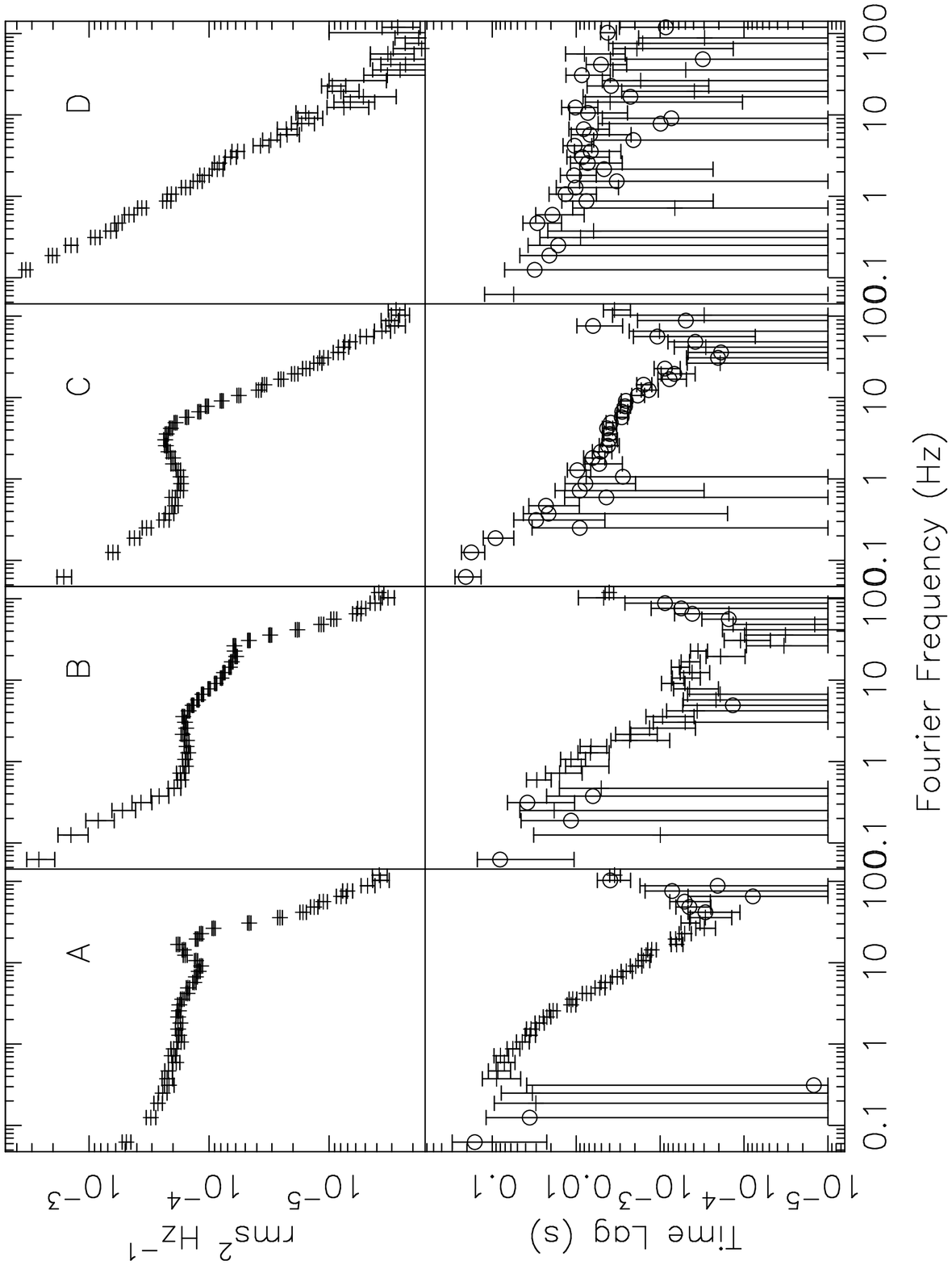]{The PDSs (upper panels) with 1.8-5.1 keV and the cross spectra
(lower panels) between 5.1-13.1 keV and 1.8-5.1 keV with 16 s segment on the vertical HB (label A), the
horizontal HB (label B), the NB (label C), and the FB (label D), respectively.
The data points outside the boxes in Fig. 1 or including gaps in the HID
are not used. The ``+" and
``$\circ$" represent the hard and soft lag respectively. The time lags evolve
from the positive to the negative with
disappearance of HBO and appearance of NBO along the Z track. There is a negative
time lag trend
below 0.3 Hz in the cross spectrum of the HB.\label{fig2}}

\figcaption[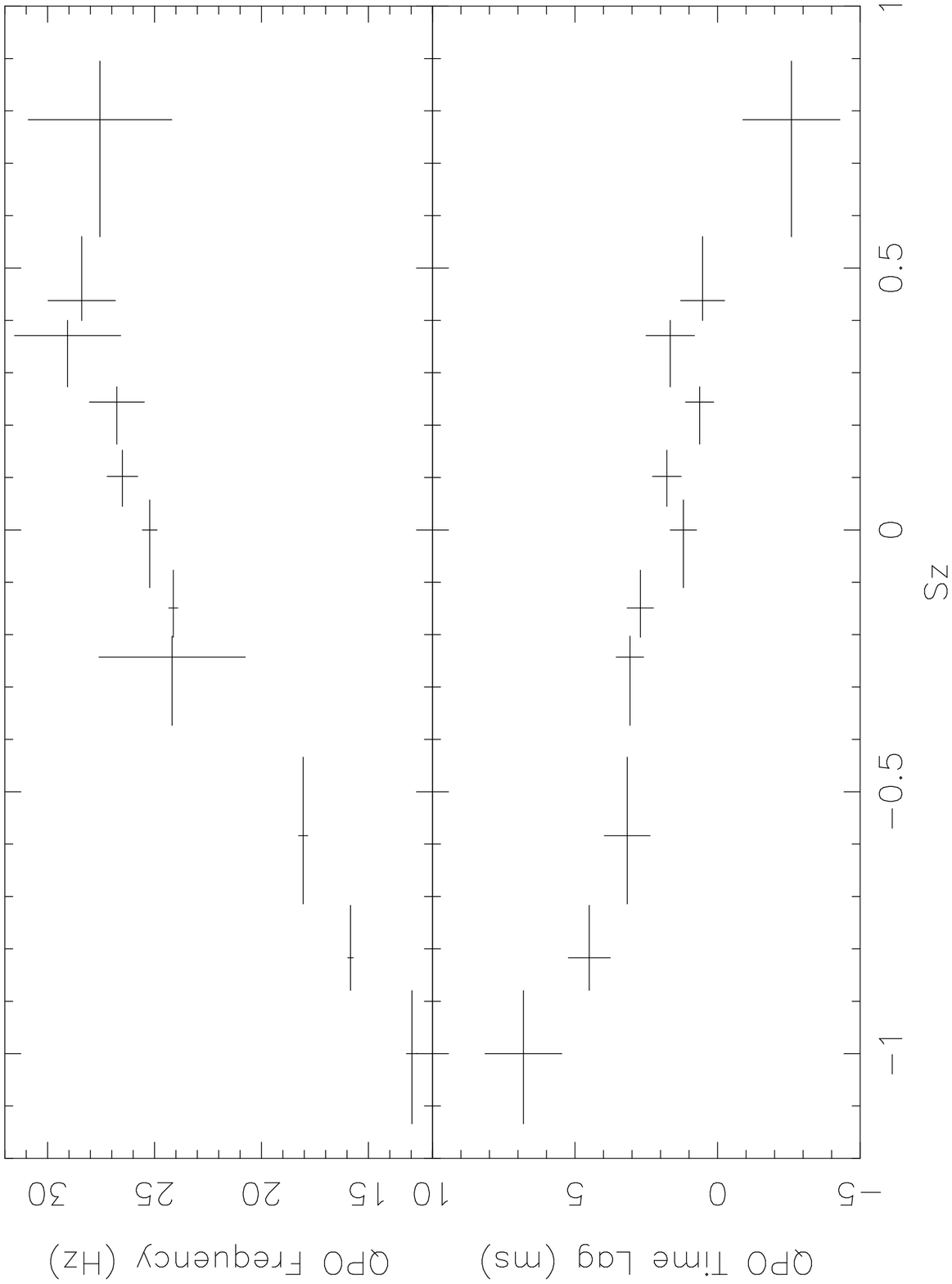]{The QPO centroid frequency (upper panel) and
the time lag of QPO (lower panel) on the HB, plotted as a function of
$S_z$,  which measures
the position of Cir~X-1 in the HID. Results from the box
11 and 12 are excluded because the QPO of these two boxes
have faded into a ``knee" and a significant centroid frequency and FWHM can not been
obtained. Note that a positive time delay is a hard lag.\label{fig3}}

\figcaption[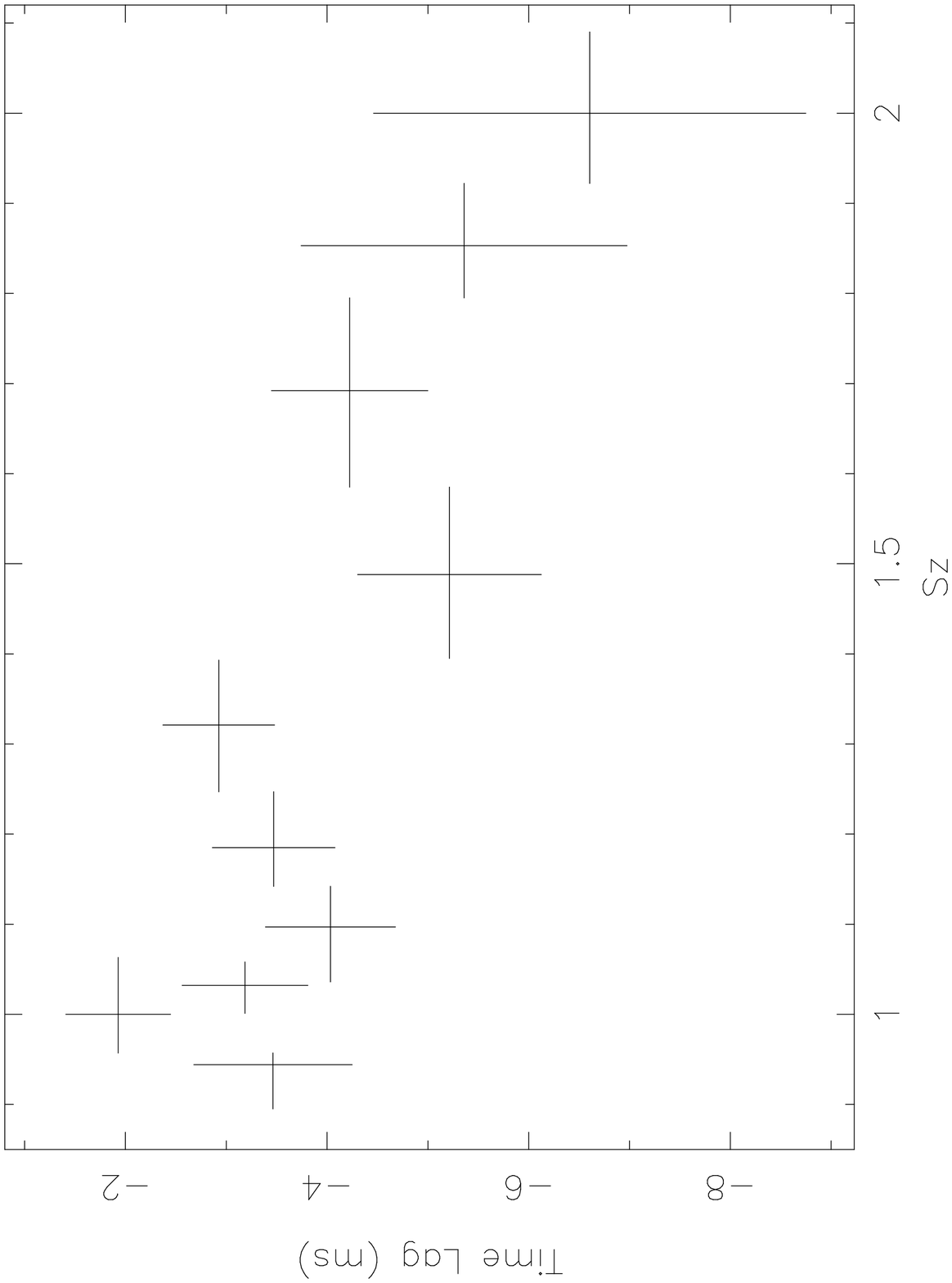]{The average time lag between 5.1-13.1 keV
and 1.8-5.1 keV in the frequency range between 2 Hz and 6 Hz on the NB and at
the end of the HB where a QPO around 4 Hz is observed.
\label{fig4}}

\clearpage

\plotone{fig1.ps}

\clearpage

\plotone{fig2.ps}

\clearpage

\plotone{fig3.ps}

\clearpage

\plotone{fig4.ps}

\end{document}